\documentclass[pra,aps,showpacs,showkeys,preprint,superscriptaddress,amsmath,amssymb,floatfix,12pt,longbibliography]{revtex4-1}
\usepackage{amsmath,amssymb,graphicx}
\usepackage[colorlinks=true, citecolor=blue, linkcolor=blue, urlcolor=blue]{hyperref}
\usepackage{array}

\begin{document}

\title{Partial Coherence in modern optics -- Emil Wolf's legacy in the 21st century}

\author{Girish S. Agarwal} \email{girish.agarwal@tamu.edu}
\affiliation{Institute for Quantum Science and Engineering, Texas A\&M University, College Station, TX 77843-4242, USA}
\affiliation{Department of Biological and Agricultural Engineering, Texas A\&M University, College Station, TX 77843-4242, USA}
\affiliation{Department of Physics and Astronomy, Texas A\&M University, College Station, TX 77843-4242, USA}

\author{Anton Classen} \email{aclassen@tamu.edu}
\affiliation{Institute for Quantum Science and Engineering, Texas A\&M University, College Station, TX 77843-4242, USA}
\affiliation{Department of Biological and Agricultural Engineering, Texas A\&M University, College Station, TX 77843-4242, USA}
\affiliation{Department of Physics and Astronomy, Texas A\&M University, College Station, TX 77843-4242, USA}

\begin{abstract}
We highlight the impact of Emil Wolf's work on coherence and polarization on an ever increasing amount of applications in the $21^\text{st}$ century. We present a brief review of how partial coherence at the level of increasing order of coherence functions is leading to evolution in the better methods for microscopy, imaging, optical coherence tomography; speckle imaging; propagation through random media. This evolution in our capabilities is expected to have wide ramifications in Science and Engineering. 
\end{abstract}


%\pacs{test}

\maketitle

\section{Introduction}

It is a great honor and privilege to contribute this article in the memory of Emil Wolf. We especially like to celebrate his contributions to the theory of partial coherence which today finds wide range of applications in Physics, Engineering and Biomedical Sciences -- some of these applications involve propagation of laser fields through turbulent atmosphere, optical image formation, medical diagnostics, optical coherence tomography, speckle imaging, superresolution microscopy, and studies of disorder. In this article we would discuss the basis for some of these applications and bring out especially the role of partial coherence which may be either due to the source used to probe or due to the statistical fluctuations in the medium. In the words of Emil Wolf \cite{Born1999}

\textit{All optical fields undergo random fluctuations. They may be small, as in the output of many lasers, or they may be appreciably larger, as in light generated by thermal sources. The underlying theory of fluctuating optical fields is known as coherence theory. An important manifestation of the fluctuations is the phenomenon of partial polarization. Actually, coherence theory deals with considerably more than fluctuations. Unlike usual treatments, it describes optical fields in terms of observable quantities and elucidates how such quantities, for example, the spectrum of light, change as light propagates.} 

Our emphasis is going to be on a whole range of observable quantities which include not only measurements of the intensities but also a whole range of intensity-intensity correlations of order two and higher. Most of our discussion would be based on classical fields and sources. In a few sections we also highlight the role of coherence and its utility in the context of quantum fields.

\section{Wolf's classic development of the theory of partical coherence}

\noindent Wolf's theory of partial coherence, as developed by Wolf in the 1950's, characterized electromagnetic radiation fields in terms of the second-order correlation function of the fields at different space-time points $(\mathbf{r}_1,t_1)$ and $(\mathbf{r}_2,t_2)$ \cite{Born1999}
\begin{equation}
\Gamma_{\alpha \beta} (\mathbf{r}_1,t_1, \mathbf{r}_2,t_2) = \Gamma_{\alpha \beta} (\mathbf{r}_1,\mathbf{r}_2,\tau) = \langle E_\alpha(\mathbf{r}_1,t+\tau) E_\beta^\ast(\mathbf{r}_2,t) \rangle \, .
\label{eq:1}
\end{equation}   
Here, the $\langle \, \ldots \, \rangle$ denote the ensemble average over the statistical fluctuations of the fields. Note that we have assumed that the fields are stationary and ergodic so that the correlation function depends only on the time difference $\tau = t_1-t_2$. The fields $E_\alpha(\mathbf{r},t)$ are analytic signals, i.e. they contain only the positive frequencies. In many applications one drops the vector character of the field and works with the scalar fields. This is allowed as long as the optical setup involves no change in the polarization properties of the field.

The time scale over which $\Gamma_{\alpha \beta} (\mathbf{r}_1,\mathbf{r}_2,\tau)$ varies significantly is called the temporal coherence $\tau_c$ of the field. The intensity $I(\mathbf{r})$ at a point can be taken as $\sum_{\alpha \alpha} \Gamma_{\alpha \alpha} (\mathbf{r},\mathbf{r},0)$, although more precisely its definition would depend on the exact experimental measurement scheme. The correlation function $\Gamma$ satisfies a set of inequalities which can be derived from the positivity of the probability distributions charaterizing the statistical fluctuations of the field(s). For instance, for a scalar field (dropping the indices $\alpha$ and $\beta$) the Schwarz inequality gives
\begin{equation}
|\Gamma (\mathbf{r}_1,\mathbf{r}_2,\tau)|^2 \leq  \Gamma (\mathbf{r}_1,\mathbf{r}_1,0) \Gamma (\mathbf{r}_2,\mathbf{r}_2,0) \, ,
\label{eq:2}
\end{equation}  
which leads to the so-called normalized coherence function
\begin{equation}
\gamma (\mathbf{r}_1,\mathbf{r}_2,\tau) = \frac{\Gamma (\mathbf{r}_1,\mathbf{r}_2,\tau)}{\sqrt{\Gamma (\mathbf{r}_1,\mathbf{r}_1,0) \Gamma (\mathbf{r}_2,\mathbf{r}_2,0)}} \; , \; |\gamma| \leq 1 \, .
\label{eq:3}
\end{equation}
Note that sometimes throughout the manuscript the temporal or spatial coordinates, $\tau$ or $(\mathbf{r}_1,\mathbf{r}_2)$, respectively, will be omitted which shall represent either equal-time or equal-point coherence/correlation functions. 

\begin{figure}[t!]%
\begin{center}
\includegraphics[width=0.6\textwidth]{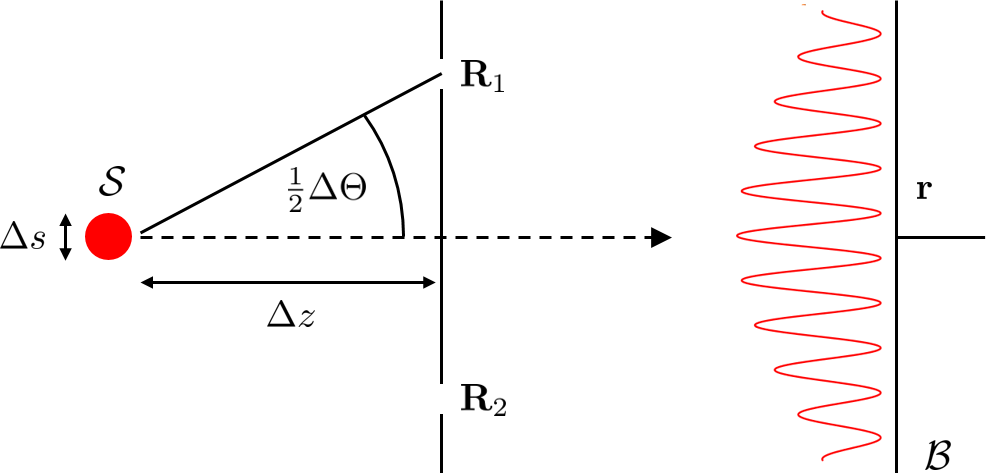}%
\caption{Young's double-slit experiment: A quasi-monochromatic source $\mathcal{S}$ with lateral width $\Delta s$ is placed in front of a double slit in a distance $\Delta z$. An interference pattern on a screen $\mathcal{B}$ is visible if the slits at $\mathbf{R}_1$ and $\mathbf{R}_2$ are located within the coherence area of the source.}%
\label{fig:Young1}%
\end{center}
\end{figure}

An electric field can be defined to be coherent (incoherent) if  $|\gamma| \rightarrow 1$ (0). The significance of $\gamma$ is especially apparent in the celebrated Young's double-slit experiment, which has been the landmark experiment in optics as it established the wave nature of light (and later also the field of quantum mechanics/optics by means of single-photon interference and single-particle interference \cite{Bach2013,Feynman1965}). The visibility of the interference fringes depends on the partial coherence of the source (and the explicit geometry of the experimental setup and the double-slit). Consider the setup shown in fig.\,\ref{fig:Young1} using a quasimonochromatic source with central frequency $\omega_l$, leading to the electric field amplitude $E(\mathbf{r},t) = \mathcal{E} (\mathbf{r},t) e^{-i \omega_l t }$, where $\mathcal{E} (\mathbf{r},t)$ is a slowly varying function of $t$. The visibility $\mathcal{V}$ at a point $\mathbf{r}$ on the screen $\mathcal{B}$ is given by 
\begin{equation}
\mathcal{V} = \frac{2 \sqrt{I_1(\mathbf{r}) I_2(\mathbf{r})}}{I_1(\mathbf{r}) + I_2(\mathbf{r})} |\gamma_{1 2}(\tau)| \, , \quad  \tau = \frac{s_2-s_1}{c} \, .
\label{eq:4}
\end{equation}  
Here $I_1(\mathbf{r})$ and $I_2(\mathbf{r})$ are the respective intensities at the position $\mathbf{r}$ if either slit one or slit two is open (while the other slit is closed). More precisely, the intensities read $I_\alpha(\mathbf{r}) \propto \langle \mathcal{E} (\mathbf{R}_\alpha,t) \mathcal{E}^\ast (\mathbf{R}_\alpha,t) \rangle$, where $\mathbf{R}_\alpha$ is the transverse coordinate in the double-slit plane. The time difference $\tau$ arises due to different optical path lengths $s_1$ and $s_2$ for light to propagate from the source $\mathcal{S}$ to the screen $\mathcal{B}$ via slit 1 and slit 2, respectively. In many cases $I_1 \sim I_2$, leading to $\mathcal{V} = |\gamma_{1 2}(\tau)|$. Thus the visibility of the interference fringes provides a direct measurement of the degree of partial coherence between the two electric fields scattered/diffracted by the two slits. In addition, if the double-slit geometry and distance is known one can also infer knowledge on the source $\mathcal{S}$.

In recent literature \cite{Eberly2017} another quantity was introduced, called distinguishability $\mathcal{D}$, which is a measure of how strong the intensities at the two slits differ
\begin{equation}
\mathcal{D} = \frac{|I_1-I_2|}{I_1+I_2} \, .
\label{eq:4a}
\end{equation} 
It can be  seen from the definitions of $\mathcal{V}$ and $\mathcal{D}$, together with the condition $|\gamma| \leq 1$ that 
\begin{equation}
\mathcal{V}^2 + \mathcal{D}^2 \leq 1 \, .
\label{eq:4b}
\end{equation}
Extensive literature exists on this inequality, especially when a Young's double-slit interference experiment is performed with quantum fields -- where $\mathcal{V}$ is associated with the wave character and $\mathcal{D}$ with the particle character.

An important consequence of the analytical result of eq.\,(\ref{eq:4}) is that if the path difference $s_2-s_1$ exceeds the coherence time $\tau_c$ of the source $\mathcal{S}$, then $\gamma \rightarrow 0$ and the interference fringes disappear. Wolf further introduced the remarkable idea of interference in the spectral domain when interferences in the time domain do not survive. In the setup of  fig.\,\ref{fig:Young1} one considered a measurement of the intensity at the point $\mathbf{r}$. However, now, instead of simply measuring the intensity at position $\mathbf{r}$, one would measure the spectrum $S(\mathbf{r},\omega)$ defined by
\begin{equation}
S(\mathbf{r},\omega) = \frac{1}{2\pi} \int_{-\infty}^{+\infty} \langle   \mathcal{E} (\mathbf{r},t+\tau) \mathcal{E}^\ast (\mathbf{r},t) \rangle e^{i\omega \tau} \text{d}\tau \, ,
\label{eq:5}
\end{equation}  
so that the intensity will be $ I(\mathbf{r}) = \int_{-\infty}^{+\infty} S(\mathbf{r},\omega) \,\text{d}\omega$. For a fixed $\tau_c$ the measured spectrum would display modulations in $\omega$. A change in $\tau_c$ changes the character of the modulations. These modulations depend on the spectral correlations between the two source points $P_1$ and $P_2$ (at $\mathbf{R}_1$ and $\mathbf{R}_2$), defined by 
\begin{equation}
S(\mathbf{R}_1,\mathbf{R}_2,\omega) = \frac{1}{2 \pi} \int_{-\infty}^{+\infty} \langle   \mathcal{E} (\mathbf{R}_1,t+\tau) \mathcal{E}^\ast (\mathbf{R}_2,t) \rangle e^{+i\omega \tau} \text{d}\tau \, .
\label{eq:6}
\end{equation}
James and Wolf demonstrated Young's interference \cite{Born1999} with broad band light from a black body source. Subsequently the idea of interference in frequency space led to a new field of research with many theoretical and experimental works \cite{James1991,Agarwal1993,Kumar2001,Zou1992} and with practical applications.

Wolf discovered that the spectrum of a partially coherent source may change upon propagation \cite{Wolf1986,Born1999}. The spectrum may show shifts, blue or red, depending on the nature of the correlations. This discovery also opened up a new field of research and surprised many who had no idea what partial coherence can lead to. Such spectral shifts were observed in a large number of experiments. An analogue of Wolf shifts was found in the microscopic problem of radiation from two trapped atoms which were separated within a wavelength of each other \cite{Varada1992}. In this case the quantum electrodynamic interaction \cite{Agarwal2012} naturally produces correlation between the two atoms; leading to spectral characteristics of the two atom system different from that of a single atom. The required correlation needed for a Wolf shift for this system is an inherent property of the quantum electrodynamic interaction.

Wolf emphasized the importance of the coherent mode decomposition of the spectral correlation function $S(\mathbf{R}_1,\mathbf{R}_2,\omega)$
\begin{equation}
S(\mathbf{R}_1,\mathbf{R}_2,\omega) = \sum_\alpha \Phi_\alpha (\mathbf{R}_1,\omega) \Phi_\alpha^\ast (\mathbf{R}_2,\omega) \lambda_\alpha \, .
\label{eq:7}
\end{equation}
The functions $\Phi_\alpha$ are the coherent modes of the system for a frequency $\omega$. The modes are the eigenfunctions of the integral equation defined over the domain of interest
\begin{equation}
\int S(\mathbf{R}_1,\mathbf{R}_2,\omega) \Phi_\alpha (\mathbf{R}_2,\omega) \text{d}\mathbf{R}_2 = \lambda_\alpha \Phi_\alpha (\mathbf{R}_1,\omega) \, .
\label{eq:8}
\end{equation}
The hermiticity and non-negative definiteness of $S$ ensures that the eigenvalues $\lambda_\alpha$ are real and non-negative. In recent years the importance of the decomposition  of eq.\,(\ref{eq:7}) is seen in the context of classical entanglement which is inherently implied by the structure of eq.\,(\ref{eq:7}) \cite{Agarwal2002,Kagalwala2013,Qian2011}

In another classic work \cite{Born1999,Wolf1959} Wolf formulated the polarization characteristics of partially coherent beam-like fields in terms of a coherence matrix with entries given by the coherence functions of the field at the same space-time point
\begin{equation}
J = \begin{pmatrix}
	\langle E_x^\ast E_x \rangle & \langle E_x^\ast E_y \rangle \\ \langle E_y^\ast E_x \rangle & \langle E_y^\ast E_y \rangle
\end{pmatrix} \, .
\label{eq:9}
\end{equation}
The degree of polarization $P$ is given by 
\begin{equation}
P = \left( 1 - \frac{4\, \text{Det}(J)}{[\text{Tr} (J)]^2} \right)^{1/2} \, , \quad 0 \leq P \leq 1 \, .
\label{eq:10}
\end{equation}
The passage of a partially coherent light field through an optical system can be formulated by a series of unitary transformations on $J$, assuming that the optical system is not lossy. He provided a unified treatment of coherence and polarization \cite{Mandel1995,Wolf2003a}. Eberly \textit{et al.} \cite{Eberly2017,Qian2018} included partial polarization in their study of the interference with polarized fields. With the inclusion of polarization the inequality of eq.\,(\ref{eq:4b}) turns into the equality
\begin{equation}
\mathcal{V}^2 + \mathcal{D}^2 = \mathcal{P}^2 \, .
\label{eq:10a}
\end{equation}
They called this equality the polarization coherence theorem \cite{Eberly2017}. This relation has strong consequences for the wave particle duality in quantum physics \cite{Qian2018,DeZela2018}

Although Wolf concentrated mostly on classical fields, all the second-order coherence functions can be generalized to quantum fields, where we write the corresponding quantum field operator as
\begin{equation}
\mathbf{E}(\mathbf{r},t) = \mathbf{E}^{(+)}(\mathbf{r},t) +  \mathbf{E}^{(-)}(\mathbf{r},t) \, ,
\label{eq:11}
\end{equation} 
where $\mathbf{E}^{(+)}(\mathbf{r},t)$ and $\mathbf{E}^{(-)}(\mathbf{r},t) $ are called the positive and negative frequency parts of electric field operator. The $\mathbf{E}^{(+)}$ part plays the same role as the analytical signal in Wolf's theory. Since $\mathbf{E}^{(+)}$ and $\mathbf{E}^{(-)}$ do not commute, the quantum theory uses ordered correlations. In most applications we measure normally ordered quantities \cite{Glauber1963-2}. Thus we replace eq.\,(\ref{eq:1}) by
\begin{equation}
\Gamma_{\alpha\beta} (\mathbf{r}_1,\mathbf{r}_2,\tau) = \langle E^{(-)}_\beta(\mathbf{r}_2,t) E^{(+)}_\alpha(\mathbf{r}_1,t+\tau) \rangle \, ,
\label{eq:12}
\end{equation}
using again scalar quantities. However, there are quantum processes like spontaneous emission which are initiated by the vacuum of the radiation field. In such cases we also need the anti-normally ordered correlations of the form
\begin{equation}
\Gamma_{\alpha\beta}^{(A)} (\mathbf{r}_1,\mathbf{r}_2,\tau) = \langle E^{(+)}_\alpha(\mathbf{r}_1,t+\tau) E^{(-)}_\beta(\mathbf{r}_2,t) \rangle \, ,
\label{eq:13}
\end{equation}
where we have added a superscript $(A)$ to indicate this anti-normal ordering. 

Wolf primarily concentrated on developments where the second-order coherence function of two electric field amplitudes was relevant. Today, however, his insights play a major role in many fields of optics. This includes a) optical coherence tomography (OCT), b) intensity-intensity correlation measurements of a (chaotic) thermal light source (TLS) as originally conducted by Hanbury Brown and Twiss (HBT) and in many modern spin-offs to measure properties of generic scattering media or non-classical two-level atoms, c) Ghost imaging with SPDC photon pairs or with thermal light sources, d) superresolution speckle illumination imaging and microscopy, also combined with the analysis of higher-order intensity correlations and cumulants, and e) superresolution fluorescence microscopy enabled by intensity correlation analyses and structured illumination.

\section{OCT based on partial coherence}

One highly applied field where partial coherence plays a major role is optical coherence tomography (OCT) \cite{Fercher2003,Park2015,Hitzenberger2018}. OCT investigates discontinuities of the refractive index and the attenuation coefficient of living tissues, mostly to identify eye diseases. 

Tomography in the x-ray and gamma ray regime relies on forward projections at different angles, i.e.~on the Fourier slice theorem to obtain depth resolution. Since optical techniques are dominated by diffraction it is not applicable there. Instead, OCT relies on the Fourier diffraction projection theorem \cite{Wolf1969}, which derives tomographic images from single back-scattered radiation. Moreover, OCT is based on interference phenomena of light waves as well as second-order amplitude (cross) correlation measurements. A full 3D image is obtained by lateral scanning of a probe beam. Compared to forward projection tomography, OCT bears some outstanding properties: 1) depth resolution is decoupled from the transverse one, 2) high depth resolution, in the histological $1\,\mu$m range, is possible, 3) high dynamic range and sensitivity, and 4) in medical terms OCT is a non-invasive technique that yields \textit{in vivo} data \cite{Fercher2003}.	

\begin{figure}[t]
  \centering
	\includegraphics[width=0.65\textwidth]{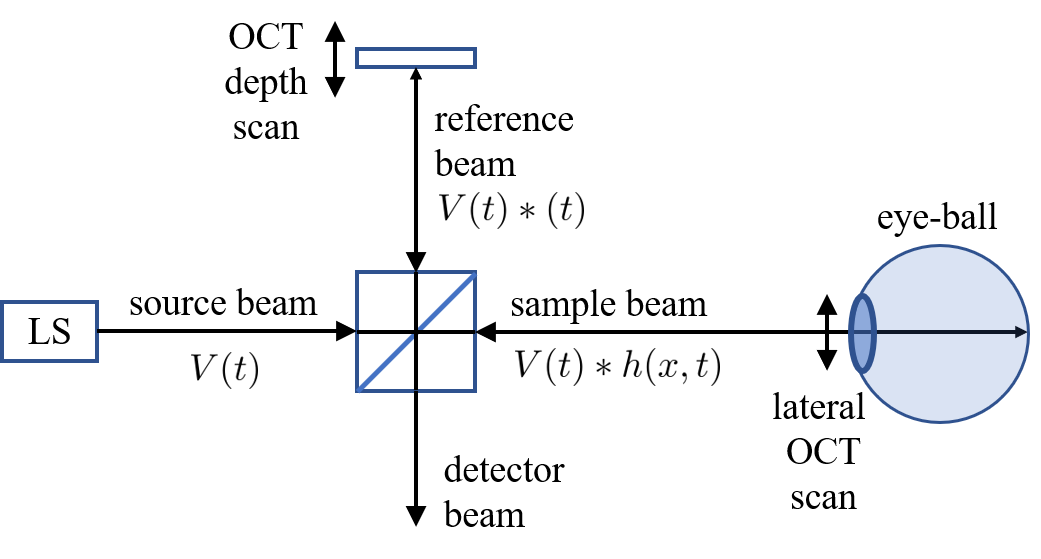}
  \caption{Standard OCT scheme based on a low time-coherence Michelson interferometer (LCI).} 
  \label{fig:oct1}
\end{figure}

Initially, OCT techniques were based on low time-coherence Michelson interferometry (LCI), where depth-scans are performed in the time domain. A schematic illustration is provided in Fig.\,\ref{fig:oct1}. The figure of merit for OCT is the second-order cross-correlation function between the sample wave amplitude $V_S$ and the reference wave amplitude $V_R$
\begin{equation}
\Gamma_{SR}(\tau) = \langle V_S^\ast(t) V_R(t+\tau)  \rangle  \, ,
\label{eq:oct1}
\end{equation}
for some time delay $\tau$. For $\tau = 0$, the average measured intensity reads $\bar{I} = \langle I(t)  \rangle = \langle V^\ast(t) V(t+\tau)  \rangle_{\tau = 0}$, with $V = V_S(t) + V_R(t+\tau)$. After introducing an adjustable time delay $\Delta t = 2 \Delta z / c$ (see Fig.\,\ref{fig:oct1}) the measured intensity becomes
\begin{equation}
I(\Delta t) = \bar{I}_S + \bar{I}_R + G_{SR}(\Delta t)
\label{eq:oct2}
\end{equation}
with the interferogram
\begin{equation}
G_{SR}(\Delta t) = 2 \sqrt{\bar{I}_S  \bar{I}_R} |\gamma_{SR}(\Delta t)| \cos(\alpha_{SR}-\delta_{SR}(\Delta t)) \, .
\label{eq:oct3}
\end{equation}
Here, $\alpha_{SR}$ is some constant phase and $\delta_{SR}(\Delta t) = 2 \pi \bar{\nu} \Delta t$. Then, to obtain the complex-valued $\Gamma_{SR}(\tau)$ of eq.~(\ref{eq:oct1}), i.e.~the sough-after OCT signal, the real-valued $G_{SR}(\Delta t)$ needs to be Hilbert transformed. In the experiment $G_{SR}(\Delta t)$ is proportional to the measured photodiode (heterodyne) signal; for further details see \cite{Fercher2003}. Finally, it is helpful to use spectral relations, which are obtained with the help of the Wiener-Khintchine theorem, leading to the spectral interference law
\begin{equation}
S(\nu,\Delta t) = S_S(\nu) + S_R(\nu) + 2 \text{Re}[W_{SR}(\nu)]\cos(2 \pi \nu \Delta t) \, ,
\label{eq:oct4}
\end{equation}
with $W_{SR}(\nu) = FT\{ \Gamma_{SR}(\Delta t) \}$ being the so-called cross-spectral density function.

Modulations of $G_{SR}(\Delta t)$ only occur due to back-scattering from sample features within a limited axial range corresponding to $\Delta t \leq \tau_c$, where $\tau_c$ is the coherence time of the utilized light source, which is assumed to have a Gaussian envelope. In time-domain OCT the depth resolution hence becomes
\begin{equation}
c \tau_c \sim \Delta z_\text{FWHM} = \frac{2 \ln(2)}{\pi} \frac{\bar{\lambda}^2}{\Delta \lambda_\text{FWHM}} \, ,
\label{eq:oct6}
\end{equation}
with the mean wavelength $\bar{\lambda}$ and the bandwidth $\Delta \lambda_\text{FWHM}$. It is thus imperative to use a broad-band low-coherence beam.

\begin{figure}[t]
  \centering
	\includegraphics[width=0.65\textwidth]{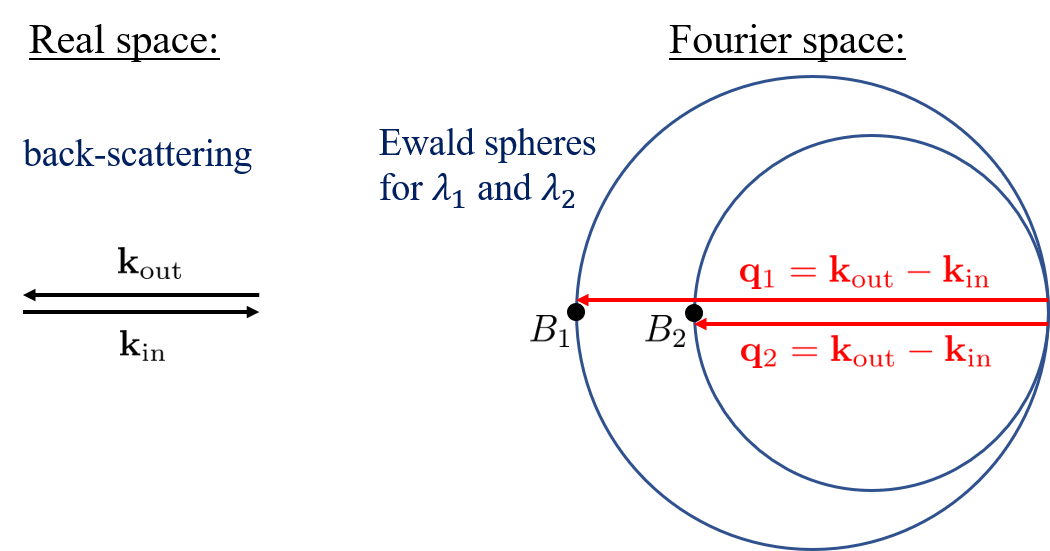}
  \caption{Ewald sphere constructions for wavelengths $\lambda_1$ and $\lambda_2$. Back-scattering in real space for the two wavelengths is represented by the two points $B_1$ and $B_2$ in Fourier space.}  
  \label{fig:oct2}
\end{figure}

Another field, comprises so-called Fourier-domain OCT techniques \cite{Fercher2003}. This includes spectral interferometry approaches. For monochromatic light of wavelength $\lambda_1$ the approach equals well-known coherent scattering methods, where a (scalar) scattering potential $V(\mathbf{R})$ produces the scattered wave
\begin{equation}
E_\text{out}(\mathbf{q}) \sim \frac{e^{ikr}}{r} \int V(\mathbf{R}) \, \exp(-i \mathbf{q} \cdot \mathbf{R}) \, E_\text{in}(\mathbf{R}) \, \text{d}^3 \mathbf{R} \, 
\label{eq:oct6a}
\end{equation}
in first-order Born approximation. Note that the illumination is often described as a simple plane wave with envelope $E_\text{in}(\mathbf{R}) = const$. The measured quantity $ E_\text{out}(\mathbf{q}) \sim f(\mathbf{q})$ represents the Fourier amplitudes of the object, with $\mathbf{q} = \mathbf{k}_\text{in} - \mathbf{k}_\text{out}$ being the momentum transfer vector. In contrast to crystallography, where the entire Ewald sphere is measured (i.e.~a 2D manifold in 3D Fourier space, see Fig.~\ref{fig:oct2}), in Fourier OCT only the back-scattered radiation with $\mathbf{k}_\text{out} = - \mathbf{k}_\text{in}$ is recorded, and thus only $f(\mathbf{q}) = f(-2\mathbf{k}_\text{in})$. To obtain useful information more axial Fourier components are required. Towards that, the process can be repeated with another wavelength $\lambda_2$ \cite{Agarwal1998}, as illustrated in Fig.~\ref{fig:oct2}. Ideally the entire range between the points $B_1$ and $B_2$ in Fourier space is accessed by use of a broad band spectrum. Experimentally this is achieved by wavelength tuning or spectral interferometry techniques. The Fourier transform of the recorded cross-spectral intensity provides the sought-after interferogram of eq.~(\ref{eq:oct3}). In contrast to time-domain techniques only the lateral OCT scanning procedure remains. The disadvantages are the neccessary detector array (camera) and the smaller dynamic range.

Other generalizations of OCT include speckle-illumination LCI, since Speckle has become a very useful phenomenon for a series of measurement techniques \cite{Sirohi1999,Briers2001}. In this case the intensity correlations (see section below for definition) can be obtained from the amplitude correlation $\Gamma(\tau)$ by \cite{Loudon2001}
\begin{equation}
\langle I(t) I(t+\tau)  \rangle = \langle I(t)  \rangle^2 [1 + |\Gamma(\tau)|^2] \, .
\label{eq:oct5}
\end{equation}
The temporal coherence length $\tau_c$ of the speckle intensity fluctuations leads to intrinsic depth resolution [similar to eq.~(\ref{eq:oct6})]. Another approach, beyond the linear optics regime, is based on SPDC and two-photon interferometry, and has shown the potential to have still higher sensitivities and two-fold enhanced depth resolutions.  Moreover, a cancellation of dispersion can be achieved \cite{Shirai2017}. Additional \textit{Functional} OCT techniques include polarization-sensitive OCT, where differences in the polarization state of the emerging light allows for the discrimination of different types of tissues \cite{Park2015}. Here, the Jones matrix formalism of eq.~(\ref{eq:9}) can be utilized. Even other techniques include Doppler OCT, enabling real-time images of \textit{in vivo} blood flow in human skin by use of the Doppler effect \cite{Zhao2002}, or a combination of Fourier and Doppler OCT which provides high phase stability and speed.

\section{Role of partial coherence in HBT-like measurements}
\label{sec:C-HBT}

In what follows we transition to the topic of (spatial) intensity-intensity correlations measurements where the fields to be correlated shall emanate from (classical) incoherent thermal light sources (TLS) such as stars. A quantum mechanical view point and correlations of light fields from non-classical sources will be discussed in section \ref{sec:Q-HBT}. In 1956, Hanbury Brown and Twiss (HBT) \cite{HBT1956-1,HBT1956-2} realized that the geometry of a single star, or even a double-star system, can be inferred from its incoherent starlight, not by means of an elaborate interferometry setup (e.g. Michelson), but simply by use of intensity-intensity correlations. While it is well-known that the (mean) intensity in the far field of an incoherent TLS geometry is a constant $I(\mathbf{r}) = I_0 = const. $, it turns out that the respective equal-time spatial second-order intensity correlations function reads 
\begin{equation}
\begin{aligned}
G^{(2)} (\mathbf{r}_1,\mathbf{r}_2, \tau = 0) & \equiv \langle  E^{(-)}(\mathbf{r}_1,t) E^{(-)}(\mathbf{r}_2,t) E^{(+)}(\mathbf{r}_2,t) E^{(+)}(\mathbf{r}_1,t)  \rangle \\
& = \langle  I(\mathbf{r}_1) I(\mathbf{r}_2) \rangle = I_0^2 + |\Gamma (\mathbf{r}_1,\mathbf{r}_2)|^2 \, .
\end{aligned}
\label{eq:hbt1}
\end{equation}  
with $\Gamma (\mathbf{r}_1,\mathbf{r}_2)$ being the second-order field correlation function of eq.\,(\ref{eq:12}). The above identification of the four-electric field product as the two-intensity product is, of course, only strictly valid for classical light fields and is also known as the Siegert relation \cite{Goodman1985}. The result can explicitly be derived via the Gaussian moment theorem \cite{Mandel1995}, since electric fields of thermal nature can be described by Gaussian random processes in phase-space with zero-mean expectation value $\langle E^{(-)}(\mathbf{r}_j)\rangle = \langle E^{(+)}(\mathbf{r}_j)\rangle = 0$. In the normalized form eq.\,(\ref{eq:hbt1}) becomes
\begin{equation}
%\begin{aligned}
g^{(2)} (\mathbf{r}_1,\mathbf{r}_2) = \frac{G^{(2)} (\mathbf{r}_1,\mathbf{r}_2)}{\Gamma (\mathbf{r}_1,\mathbf{r}_1) \Gamma (\mathbf{r}_2,\mathbf{r}_2)} \, , \qquad g^{(2)} (\mathbf{r}_1,\mathbf{r}_2) = 1 + |\gamma (\mathbf{r}_1,\mathbf{r}_2)|^2 \, ,
%\end{aligned}
\label{eq:hbt2}
\end{equation}  
For TLS $\gamma$ can be calculated by means of the van Cittert-Zernicke theorem to be the (normalized) Fourier transform of the incoherent intensity distribution $I(\mathbf{R})$ in the object plane 
\begin{equation}
  \gamma (\mathbf{r}_1,\mathbf{r}_2) = \frac{\int I(\mathbf{R}) e^{-i k [(\mathbf{u}_2-\mathbf{u}_1)\mathbf{R}]} \text{d}^2 \mathbf{R}}{\int I(\mathbf{R}) \text{d}^2 \mathbf{R}} \, ,
	\label{vcz}
\end{equation}    
where $\mathbf{u}_j$ $(j=1,2)$ denotes the unit vector pointing from the origin to the point $\mathbf{r}_j$.

For an incoherent light field $|\gamma(\mathbf{r},\mathbf{r})| = 1$ such that structural information carried by the electric fields is not preserved when measuring the (mean) intensity across the detection plane. By contrast, $g^{(2)} (\mathbf{r}_1,\mathbf{r}_2)$ accesses the amplitude information of $\gamma (\mathbf{r}_1,\mathbf{r}_2)$, while the phase information is lost. Aside from the offset of +1, the result is equivalent to coherent diffraction when measuring the intensity $I(\mathbf{q}) \sim |E_\text{out}(\mathbf{q})|^2 \sim |f(\mathbf{q})|^2$. While the amplitude information of $f(\mathbf{q})$ is accessed, the phase information which is equally required for the inversion of eq.~(\ref{eq:oct6a}) is lost. A fact and obstacle that is well-known in x-ray crystallography as the phase problem, and was taken up by Wolf in his later years \cite{Wolf2009}.     

In recent years the field of HBT intensity interferometry has been generalized to the measurement of higher-order intensity correlations. For instance, it has been predicted that superresolving imaging of equidistant 1D arrays of TLS, but later also arbitrary 2D TLS geometries, can be achieved through the analysis of higher-order correlations \cite{Thiel2007,Oppel2012a,Classen2016,Schneider2017}, both in the optical and in the x-ray regime \cite{Classen2017b, Schneider2017}. Moreover, HBT-like intensity correlation measurements have been connected to the phenomenon of Dicke super- and subradiance \cite{Wiegner2011a,Oppel2014a,Wiegner2015b,Bhatti2016,Bhatti2018a}, and even to the generation of $N00N$-like interferences from two TLS \cite{Bhatti2018b}. Other approaches consider the measurement of triplet correlations to gain direct access to phase information (at least to a certain degree) \cite{Dravins2013,Malvimat2014}. Moreover, correlation analysis has become a widespread tool in various fields of physics, ranging from stellar interferometry to nuclear collisions \cite{Baym1998,Padula2005}.

Aside from spatial correlations, the analysis of temporal correlations is of equal interest [see, for instance, eq.~(\ref{eq:oct5})]. The temporal intensity correlation function for quantum fields reads 
\begin{equation}
%\begin{aligned}
G^{(2)} (\tau )  = \langle  E^{(-)}(t) E^{(-)}(t+\tau) E^{(+)}(t+\tau) E^{(+)}(t)  \rangle %\langle  I(t) I(t+\tau) \rangle 
\approx \langle I(t)  \rangle^2 [1 + |\Gamma(\tau)|^2] \, ,
%\end{aligned}
\label{eq:hbt2a}
\end{equation} 
where the last transformation is again valid only for chaotic light fields and $\Gamma(\tau)$ is defined by eq.~(\ref{eq:12}). The normalized version $g^{(2)} (\tau)$ reports on the photon statistics of a classical or quantum light source, and more precisely on it's temporal variance/fluctuations. For a perfect laser $g^{(2)} (\tau) = 1$, meaning the photon stream is completely uncorrelated at all times. For a TLS $g^{(2)} (\tau \ll \tau_c) = 2$, with $\tau_c$ being the second-order coherence time of the light field. A TLS thus expresses photon bunching, that is a two-fold enhanced probability for a photon to arrive on the detector right after the previous one. For large time differences, however, the photons are uncorrelated $g^{(2)} (\tau \gg \tau_c) = 1$. Beyond the realm of classical optics $g^{(2)} (\tau) < 1$ can be found. This peculiar quantum feature is known as antibunching and possesses no classical analogue. The light from a single two-level atom fulfills $g^{(2)} (\tau < \tau_c) < 1$ since the atom can only emit a single photon at once. The most general approach would be to look at the spatio-temporal correlation function  $g^{(2)} (\mathbf{r}_1,\mathbf{r}_2,\tau) = 1 + |\gamma (\mathbf{r}_1,\mathbf{r}_2,\tau)|^2$ \cite{Born1999}. This $g^{(2)}$ function has a simple structure for Gaussian fluctuations and hence precise measurements of deviations can provide information on the non-Gaussian nature of fluctuations. The full non-Gaussian character can, however, be revealed only by the study of higher-order correlations.

\begin{figure}[t]
  \centering
	\includegraphics[width=0.7\textwidth]{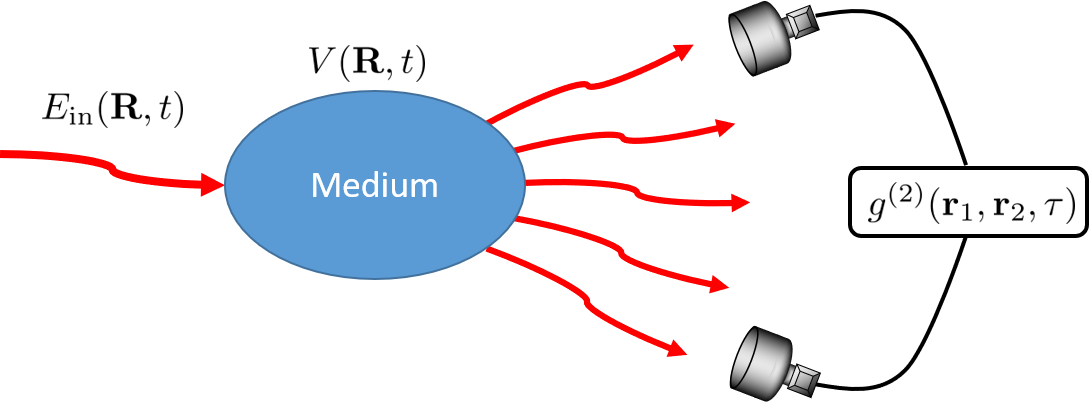}
  \caption{Schematic of light scattering by a generic medium or sample under investigation. The information about the medium is obtained from the study of the intensity and the intensity correlations of different orders.} 
  \label{fig:medium}
\end{figure}

Intensity interferometry measurements also form the basis for a plethora of methods that investigate properties of generic media that can be described as a scattering potential $V = V(\mathbf{R},t)$ in eq.~(\ref{eq:oct6a}). A schematic illustration of the idea is given in Fig.~\ref{fig:medium}. A (coherent) light field illuminates/probes an unknown medium. The light fields are modified and scattered, and afterward evaluated by means of $g^{(2)}(\mathbf{r}_1,\mathbf{r}_2,\tau)$  measurements. Given previous assumptions, the functional form of $g^{(2)}$ allows to infer various attributes of the media. Additionally the spatial and/or temporal coherence of the input probe field can be varied to study scattering from isolated areas, i.e. assuming $ E_\text{in} = E_\text{in}(\mathbf{R},t)$ in eq.~(\ref{eq:oct6a}). A list of common techniques includes dynamic light scattering (DLS), x-ray photon correlation spectroscopy (XPCS), fluorescence correlation spectroscopy (FCS), laser speckle correlation imaging (LSCI), time-domain diffuse correlation spectroscopy (TD-DCS), low-coherence enhanced back-scattering (LEBS) \cite{Young2006}, imaging through turbid media \cite{Katz2014,Smith2018}, and many more. For instance, $g^{(2)}$ can provide information on the dynamics of density fluctuations within ultracold atom clouds \cite{Shafi2015}. All these works and methods clearly demonstrate the distinct advantage of using partial spatial coherence of the source.

\section{HBT as two-photon interference and correlations of light fields from non-classical sources}
\label{sec:Q-HBT}

The results from section \ref{sec:C-HBT} can equally be interpreted in terms of two- and multi-photon interference within the quantum path formalism \cite{Fano1961,Feynman1965,Liu2009}. While interference of light fields from incoherent TLS can be modeled by classical wave theory alone,  this approach enables the exact calculation of correlation functions for non-classical light fields. The resulting expressions and modulations can be interpreted as superpositions of i) different and distinguishable multi-photon quantum paths, or ii) different, yet indistinguishable multi-photon quantum paths, weighted by the applicable quantum statistics.

\begin{figure}[t]
  \centering
	\includegraphics[width=0.65\textwidth]{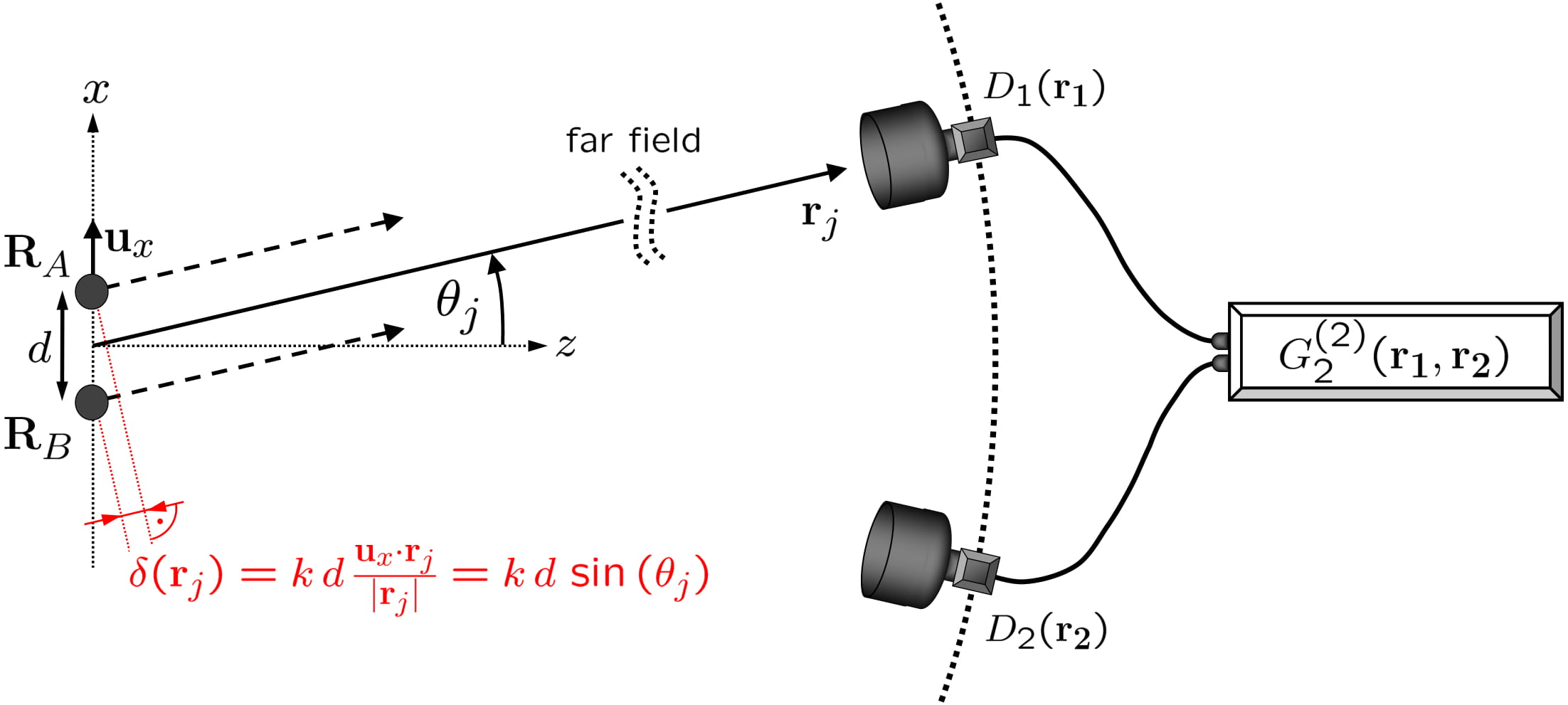}
  \caption{Two-photon coincidence detection scheme for $N=2$ point-like sources. The sources $A$ and $B$ are located at positions $\mathbf{R}_l$ $(l =A,B)$ along the $x-$axis with separation $d$. $m=2$ detectors $D_j$ located at $\mathbf{r}_j$ $(j =1,2)$ in the $x-z$ plane measure the second-order correlation function.} 
  \label{fig:setup_g2_2}
\end{figure}

\begin{figure}[t!]
  \centering
	\includegraphics[width=0.9\textwidth]{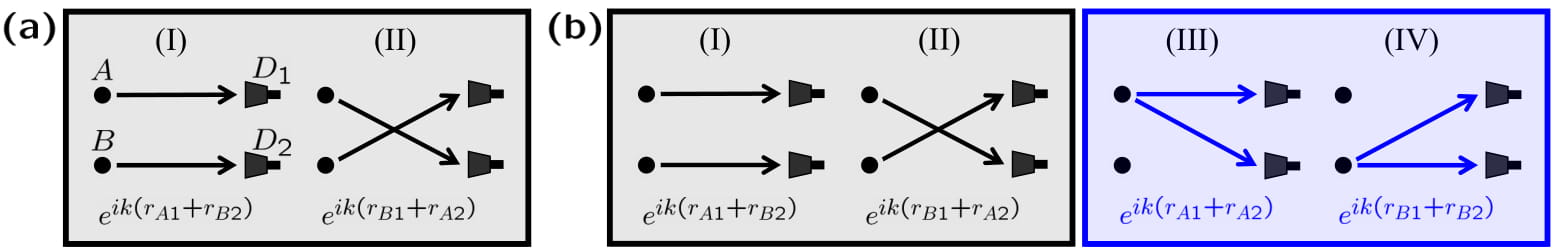}
  \caption{Two-photon quantum paths for two statistically independent light sources and two detectors. (a) shows the possible quantum paths for two independent atoms denoted by (I) and (II). For two TLS shown in (b) two additional quantum paths (III) and (IV) have to be taken into account, which lead to a constant offset.}
  \label{fig:paths_g2}
\end{figure}

To elucidate the approach let us consider the paradigmatic setup depicted in Fig.\,\ref{fig:setup_g2_2}, where two detectors in the far field measure the intensity-correlation function of two point-like emitters. The valid two-photon quantum paths leading to a coincident detection event are depicted in Fig.\,\ref{fig:paths_g2}. If source $A$ and $B$ each emit a single photon, there are two possible, yet indistinguishable two-photon quantum paths [(I) and (II) in Fig.~\ref{fig:paths_g2}(a)] for the two-photon quantum state to propagate from the two sources in the object plane to the two detectors in the far field, to induce a coincident click event. For two atoms these are the only valid quantum paths. For classical light sources the additional quantum paths (III) and (IV) exist, where either source $A$ emitted two photons or source $B$ emitted two photons, respectively [see the blue box in Fig.~\ref{fig:paths_g2}(b)]. Since the initial and final states are distinguishable here they merely add a constant offset to the signal [see e.g.~eq.~(\ref{eq:hbt2})]  


Within the quantum mechanical formalism, the ratio of spontaneous emission by two-level atoms in the excited state is directly determined by the correlation function of eq.\,(\ref{eq:13}). Coupling the atomic raising $s^{+}_i$ and lowering $s^{-}_i$ operators of each atom at Position $\mathbf{R}_i$ to the photonic annihilation and creation operators, respectively, $\gamma$ becomes \cite{Agarwal2012} 
\begin{equation}
\begin{aligned}
\gamma(\mathbf{r}_1,\mathbf{r}_2) = \frac{1}{\langle I(\mathbf{r},t) \rangle} \sum_{i,j} \langle s^{+}_i s^{-}_j  \rangle e^{i k (\mathbf{n}_1 \mathbf{R}_i -  \mathbf{n}_2 \mathbf{R}_j)} = \frac{1}{I_0} \sum_{i} \langle s^{+}_i s^{-}_i  \rangle e^{i k (\mathbf{n}_1 -  \mathbf{n}_2 ) \mathbf{R}_i} \, , 
\end{aligned}
\label{eq:14}
\end{equation}
where $\mathbf{n}_1$ and $\mathbf{n}_2$ again are the unit vectors pointing toward the directions of the detectors in the far field. Note that we assumed no dipole-dipole interactions between the atoms, nor any other coherences $\langle s^{+}_i s^{-}_j  \rangle \neq 0$, induced e.g. by coherent driving \cite{Skornia2001a,Skornia2001b,Skornia2002}. The $\langle n_i  \rangle = \langle s^{+}_i s^{-}_i  \rangle \leq 1$ denote the average population per atom, and $\mathbf{q} \equiv k (\mathbf{n}_1 -  \mathbf{n}_2 ) $ can be identified as the momentum transfer vector (cf.~Fig.~\ref{fig:oct2}).

In section \ref{sec:C-HBT} we learned that for classical light fields the $g^{(2)}$ function relates to $\gamma$ via the Siegert relation of eq.\,(\ref{eq:hbt2}), and thus to the TLS geometry encoded in the (classical) interference patterns. Recently it was shown that a very similar relationship holds true for the $g^{(2)}(\mathbf{r}_1,\mathbf{r}_2)$ function of a 3D ensemble of $N$ equal but independent single two-level atoms. It can be expressed in terms of $\gamma(\mathbf{r}_1,\mathbf{r}_2)$ as \cite{Classen2017b}
\begin{equation}
g^{(2)}(\mathbf{r}_1,\mathbf{r}_2) = 1 - \frac{2}{N} + |\gamma(\mathbf{r}_1,\mathbf{r}_2)|^2 \, .
\label{eq:hbt3}
\end{equation} 
The difference compared to eq.\,(\ref{eq:hbt2}) is the $-2/N$ summand, which arises from the non-classical statistics of atoms. Note that for $N=1$, $|\gamma|=1$ and the equal-time $g^{(2)}$ function is zero as expected from the antibunching property for a single atom. For large $N$, $g^{(2)}$ becomes equivalent to a thermal source which is reminiscent of results from the central limit theorem of probability theory. Considering an array of independent lasers with Poisson statistics would yield the summand $-1/N$. The different summands are the result of the different moments $\langle \hat{a}_i^{\dagger}  \hat{a}_i^{\dagger}  \hat{a}_i \hat{a}_i \rangle_\text{atom} = 0 $,  $\langle  \hat{a}_i^\dagger   \hat{a}_i^\dagger  \hat{a}_i \hat{a}_i \rangle_\text{Coh} = \bar{n}_i^2$ and  $\langle  \hat{a}_i^\dagger   \hat{a}_i^\dagger  \hat{a}_i \hat{a}_i \rangle_\text{TLS} = 2! \bar{n}_i^2$.  The structural information from the intensity correlation function is however contained within the \textit{classical} $\gamma$. In some sense, it can thus be argued that Wolf's theory of partial coherence already anticipated results of two-photon interference, while treating all electric fields fully classical. 

\section{Ghost imaging with thermal sources}

\begin{figure}[t!]
  \centering
	\includegraphics[width=1.0\textwidth]{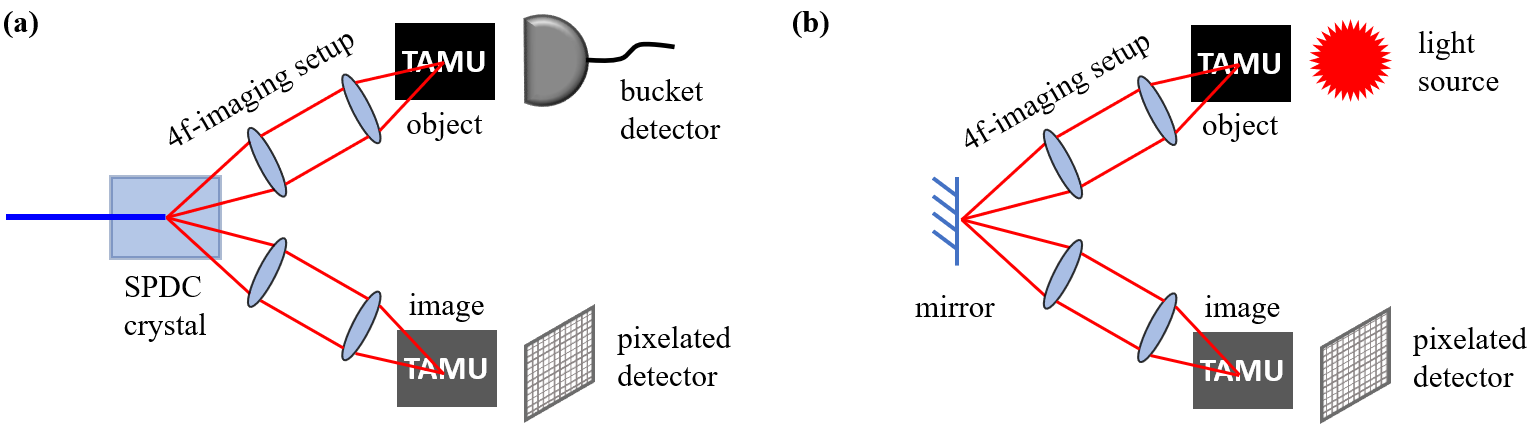}
  \caption{The images produced by a GI system (a) based on spontaneous parametric down-conversion (SPDC) are
equivalent to those that could be produced by a classical imaging system (b), albeit the GI system has a different
time sequence of events. Schematic adopted from \cite{Padgett2017}.}
  \label{fig:GI-1}
\end{figure}

Ghost imaging (GI) is a technique that makes use of spatially correlated twin beams. It correlates the outputs from two photodetectors to form the image of an unknown object. Notably, the imaging resolution is provided by the beam that never interacted with the object. A schematic illustration is given in Fig.~\ref{fig:GI-1}(a). The first step is to create spatially correlated beams or photon pairs. In the original approach a non-linear crystal for spontaneous parametric down-conversion (SPDC) was utilized \cite{Strekalov1995,Pittman1995}. The object beam illuminates the object, while a bucket detector collects the entire light that passes through. The imaging beam (which is correlated to the object beam), is measured by a high-spatial-resolution detector, i.e. a scanning pinhole or a pixelated CCD camera \cite{Erkmen2010}. Evaluating the spatial intensity correlation function between the two signals yields the image of the object. Initially assumed to be purely quantum, it was later realized that GI can equally be conducted with classically correlated twin beams. As such, GI with a chaotic thermal light field, split into two identical copies, was presented \cite{Cheng2004,Valencia2005,Gatti2008}. Aside from an offset, the measured correlation functions provide the same results. The imaging resolution and visibility are determined by the size of the speckle grains in the object plane and its feature complexity.

In other variants, such as Fourier-plane GI, the measured correlation might also represent the Fourier spectrum of the object. To understand the GI image formation process in a simple way it is helpful to consider the schematic illustration Fig.~\ref{fig:GI-1}(b). Replace the bucket detector by a light source that illuminates the object from behind. Light propagating through or scattered by the object reaches the crystal surface which now acts as a simple mirror. The reflected light propagates through the imaging arm, where it might be manipulated by lenses. Finally the intensity is measured on the CCD camera, in lieu of the intensity correlation. Simple principles of light propagation and Fourier optics allow to calculate the image on the camera. The crucial aspect is to understand that here the mirror mimics the position and momentum correlation between the twin beams. 

Another variant, known as computational GI, removes the need for the imaging arm and the spatially resolving CCD detector. Here, instead of using chaotically fluctuating twin beams, the phase front of a (coherent) beam is willingly and knowingly modified by a spatial light modulator (SLD) into many (orthogonal) variations. Since the action of the SLD on the light field is known, there is no need to monitor it with the CCD camera in the imaging arm. A variation of this approach, which inverts the position of the light source and the bucket detector (and thus the flow of light) is called single-pixel camera \cite{Padgett2017}. Very recently optical GI techniques were also taken up by the x-ray community \cite{Pelliccia2016,Yu2016}.

Structured coherence or illumination is another means to enhance the performance. In a typical Fourier GI setup the Fourier transform magnitudes are measured. The phase information is however lost. Recently a tomographic procedure was introduced in which the correlation properties of a medium were retrieved by illuminating it with an incident beam of adjustable coherence properties while recording the scattered intensity \cite{Baleine2004a,Baleine2004b}. Using a similar approach in Fourier GI with chaotic thermal light, it was shown that both amplitude and phase information can be accessed \cite{Baleine2006a}. Let us consider a modulated intensity pattern across the incoherent TLS [i.e. across the mirror plane in Fig.~\ref{fig:GI-1}(b)]
\begin{equation}
I_S(x) \sim  1 + m \cos(k x/f  \cdot \mathbf{l}_m - \phi) \, ,
\label{eq:2006-1}
\end{equation}
with modulation visibility $m$, and an adjustable phase term $\phi$. The fringe spacing is determined by the vector $\mathbf{l}_m$.  The coherence function then becomes 
\begin{equation}
\Gamma(\mathbf{r},\mathbf{r}') = g(\Delta \mathbf{r}) + \frac{m}{2} \exp(i \phi) g(\Delta \mathbf{r} - \mathbf{l}_m) + \frac{m}{2} \exp(- i \phi) g(\Delta \mathbf{r} + \mathbf{l}_m) \, ,
\label{eq:2006-2}
\end{equation}
where $\Delta \mathbf{r} = \mathbf{r} - \mathbf{r}'$ and $g(\Delta \mathbf{r})$ is an envelope function that relates to the transverse coherence length. i.e. the speckle size. The coherence function $\Gamma$ leads to a $G^{(2)}$ correlation function that carries the sought-after information. Measuring it for different phases $\phi = -\pi/2, 0, \pi/2, \pi$ creates a linear system of equations which allows to extract said information. This technique is particularly attractive for x-ray diffraction because of the incoherence of most x-ray sources and the difficulty of fabricating lenses for such short wavelengths \cite{Baleine2006a}. 

\section{Speckle illumination imaging -- advantages of intensity-intensity correlations}

\begin{figure}[t]
  \centering
	\includegraphics[width=0.73\textwidth]{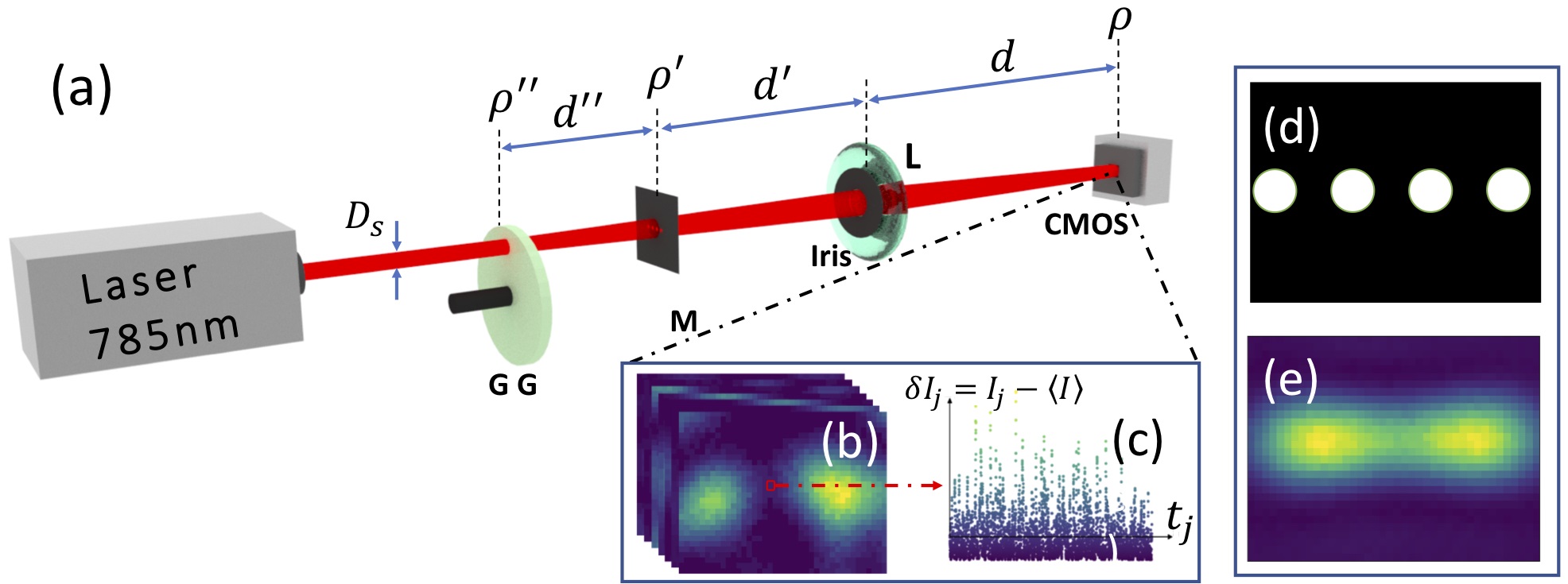}
\caption{The experimental setup.  (a) The laser incident on a rotating ground glass (GG) generates speckled light. The light transmitting through the mask (M) and iris is collected by the camera (CMOS) with a lens (L) f=150mm. Here, distance $d'=d=300mm$. (b) Several image frames are recorded to compute the high order correlation images. (c) A pixel's temporal intensity fluctuation.  (d) The mask object, notably four dots. (e) Laser illumination intensity (Int.) imaging without Ground Glass (GG), for which the image of the four dots is blurred since the Rayleigh limit is twice the distance of the dot separation \cite{Li2019a}.} 
  \label{fig:fu11}
\end{figure}

One interesting approach of using partially coherent light is speckle illumination for (coherent) imaging and microscopy \cite{Oh2013,Li2019a}. A coherent laser is spatially and temporally randomized (see Fig.\,\ref{fig:fu11}) by use of a rotating ground glass (GG) or by propagation through a random turbid medium. The speed of temporal variation determines the temporal coherence time, while the laser spot size on the GG, and other experimental parameters, determine the speckle grain size of the spatially randomized field at the plane of the object. The conjunction of both effects creates a so-called pseudo thermal light field.

Using a plane-wave coherent (laser) illumination and a lens-based imaging system to image the features of a transmission object, the resolution is limited by diffraction to around $\Delta x \approx \lambda / \mathcal{A}$, with $\mathcal{A}$ the numerical aperture of the imaging system. Note that there is approximately a factor of two difference compared to the well-known Rayleigh limit $\Delta x \approx 0.61 \, \lambda / \mathcal{A}$  for incoherent/fluorescence imaging, which arises from the fact that for coherent electric fields the imaging point-spread-function (PSF) is given by $h_\text{el}(x) = 2 J_1(x)/x$, whereas the intensity PSF for intrinsically incoherent signals reads $h_\text{int}(x) = (2 J_1(x)/x)^2$. 

\begin{figure}[t!]
  \centering
	\includegraphics[width=0.7\textwidth]{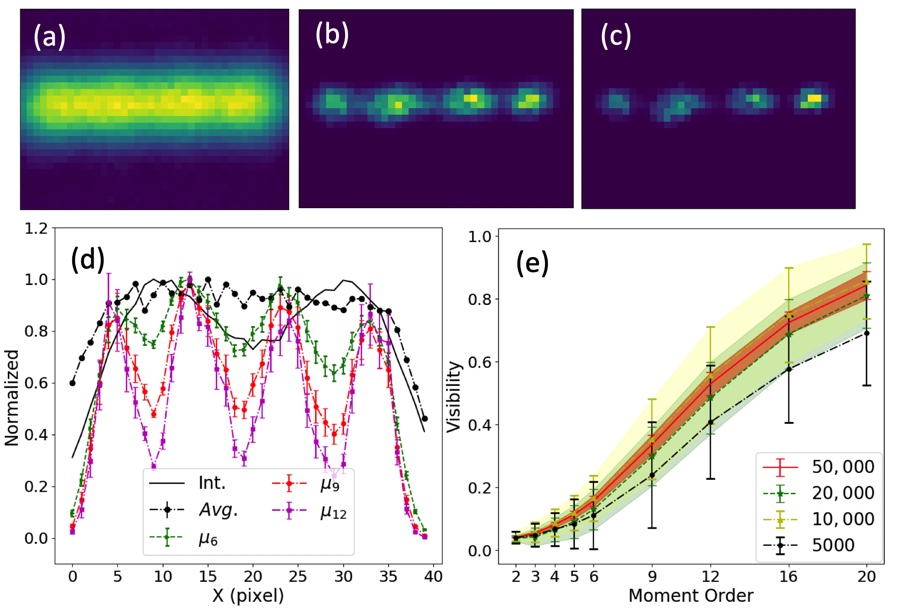}
  \caption{Comparison of traditional intensity imaging and high order moment imaging. (a) The average intensity (Avg.) imaging of speckle illumination, and the images reconstructed by the $12^{th}$, $20^{th}$ order central moment (b, c). (d) The contrast comparison from images of different orders by summing two rows of pixels that are centered with the holes. (f) The visibility and standard deviation as a function of the moment orders computed by using the different frame numbers \cite{Li2019a}.}
  \label{fig:fu22}
\end{figure}

Mutual incoherence of object features thus already provides a resolution enhancement over coherent imaging. Speckle illumination with a small grain size, i.e. with a small transverse coherence length, achieves this goal, as has recently been demonstrated by Oh \textit{et al.} \cite{Oh2013}. See also the different resolutions in the images Fig.\,\ref{fig:fu11}(e) and Fig.\,\ref{fig:fu22}(a). The effect can further be boosted via an (image plane) intensity auto-correlation analysis of a temporal sequence of such speckle illumination patterns. The fundamental principle behind this additional boost lies in the fact that super-Poissonian bunching statistics only occurs within the limited range of one speckle grain size. Positive correlations between different areas of the speckle field do not exist. After some computational post-processing \cite{Oh2013,Li2019a} the new effective PSF becomes $h_\text{eff}(x) \approx h_\text{int}(x)^2 = (2 J_1(x)/x)^4$, which equals a direct reduction of the FWHM of the PSF by a factor of $\sqrt{2}$. Including deconvolution, i.e. making use of the full spatial frequency spectrum carried by $h_\text{eff}(x)$, the resolution can be enhanced two-fold in total.  

Second-order correlations analysis, as already extensively used in imaging applications, moderately enhances the resolution. A theoretical analysis conducted by Fu Li \textit{et al.} \cite{Li2019a} shows, however, that significantly improved imaging resolutions can be achieved via the study of increasingly higher order intensity correlations. Especially the evaluation of so-called cumulants of the speckle field illumination provides enhancements well beyond the Rayleigh limit. Recently experimental evidence was provided with up to $20^\text{th}$-order correlations and cumulants. The experimental details are depicted in depicted schematically in Fig.\,\ref{fig:fu11}, where an opaque mask with four circular holes was imaged by use of an lens-based imaging system. For coherent illumination the holes are not resolved and the entire structure is skewed [see Fig.\,\ref{fig:fu11}(e)]. Utilizing speckle illumination and (average) intensity measurement leads to the better resolved structure of Fig.\,\ref{fig:fu22}(a). Nonetheless the individual features are still smeared out beyond recognition. Only when utilizing the higher-order correlations [Fig.\,\ref{fig:fu22}(b) and (c)] the structure of hole mask is well-resolved.

An interesting extension of this method would be in the imaging of gray objects. The results indicate the capability of higher-order intensity cumulants in super-resolution applications where speckles are used.  This method widens the possibilities for high order correlation imaging specifically for uses in bio-imaging and astronomy. In biomedical optics, one of standard imaging methods is laser speckle contrast imaging (LSCI), which is based on the $2^\text{nd}$-order correlation \cite{Boas2010,Davis2016,Briers1996,Aminfar2019a,Zhang2019a}

Another interesting approach of using speckle illumination along with intensity correlations can be found in fluorescence microscopy. Here, Cho et al. \cite{Kim2015} showed that superresolution microscopy can be achieved with a practical enhancement factor of 1.6 over conventional widefield microscopy (when including deconvolution). Note that the difference to the approaches described in the above paragraphs lies in the incoherent response of fluorophores to a given illumination. The varying speckle illumination thus induces chaotic and independent intensity fluctuations of the fluorescence emission of fluorophores separated by more than one speckle size. 

\section{Superresolution microscopy via intensity correlations and structured illumination}

The previous section described speckle illumination to introduce random and independent fluctuations for imaging purposes. Yet,  intrinsic properties of the fluorophores can equally be utilized for superresolution microscopy. For these intensity correlation microscopy (ICM) techniques, statistically blinking fluorophores \cite{Dertinger2009} or quantum emitters that exhibit anti-bunching \cite{Schwartz2013} can be used to enhance the resolution. Especially the first approach, known as super-resolution optical fluctuation imaging (SOFI) \cite{Dertinger2009}, is widely used in microscopy. In mathematical terms the explicit form of the (final) ICM signal, considering $m^\text{th}$-order correlations, reads 
\begin{equation}
\text{ICM}_m(\mathbf{r}) = \sum_{i=1}^N h(\mathbf{r}-\mathbf{r}_i)^m \, ,
\label{eq:icm1}
\end{equation} 
where the effective PSF becomes $h_\text{eff}(\mathbf{r}) = h(\mathbf{r})^m$. That is, the original PSF is taken to the $m^\text{th}$ power  and thus directly shrunk by the factor $\sqrt{m}$. Including deconvolution the resolution can be enhanced up to $m$-fold.

Another important application of structured coherence fields is (superresolution) structured illumination microscopy (SIM) \cite{Gustafsson2000,Heintzmann1999}. Here, a dense illumination pattern of the form $I_\text{str}(x) = 1 + \cos(k_0x + \varphi)$ illuminates an object stained with fluorophores, where $k_0$ is spatial frequency of the standing wave pattern and $\varphi$ is an adjustable phase. The fluorescence response imaged onto the detector is thus modulated by this pattern. Due to the moir\'{e} effect, the vector $k_0$ mixes with the object's spatial frequencies and encodes information from outside the original OTF support, defined by $H(\mathbf{k}) = FT\{ h(\mathbf{r})\}$. Taking a set of linearly independent images and post-processing allows for the retrieval of this information. The value $k_0$ should be maximized to reach the highest resolution and can reach $k_0=k_\text{max}$ within linear wave optics, thus enabling only a two-fold enhanced resolution. Nonetheless, the SIM toolbox is considered one of the most powerful and versatile superresolution techniques, due to its combination of resolution improvement with good acquisition speed and flexibility of use \cite{Strohl2016}. The mathematical description of SIM reads
\begin{equation}
%\begin{aligned}
\text{SIM}(\mathbf{r})  = h(\mathbf{r}) \ast \left[ n(\mathbf{r})  \times I_{\text{str}}(\mathbf{r},\alpha,\varphi) \right] = \sum_{i=1}^N  h(\mathbf{r}-\mathbf{r}_i)  \times I_{\text{str}}(\mathbf{r}_i,\alpha,\varphi) \, 
\, .
%\end{aligned}
\label{eq:sim1}
\end{equation}
with the fluorophore distribution $n(\mathbf{r})$ (which here is assumed to be a sum of point-like delta peaks) and the orientation of the structured pattern $\alpha$. Saturated SIM enhances the resolution further by introducing higher harmonics of $\cos(k_0x)$ trough a non-linear fluorophore response, though at the cost of requiring high intensities \cite{Gustafsson2005}. 

\begin{figure}[t]%
\centering
\includegraphics[width=0.8 \linewidth]{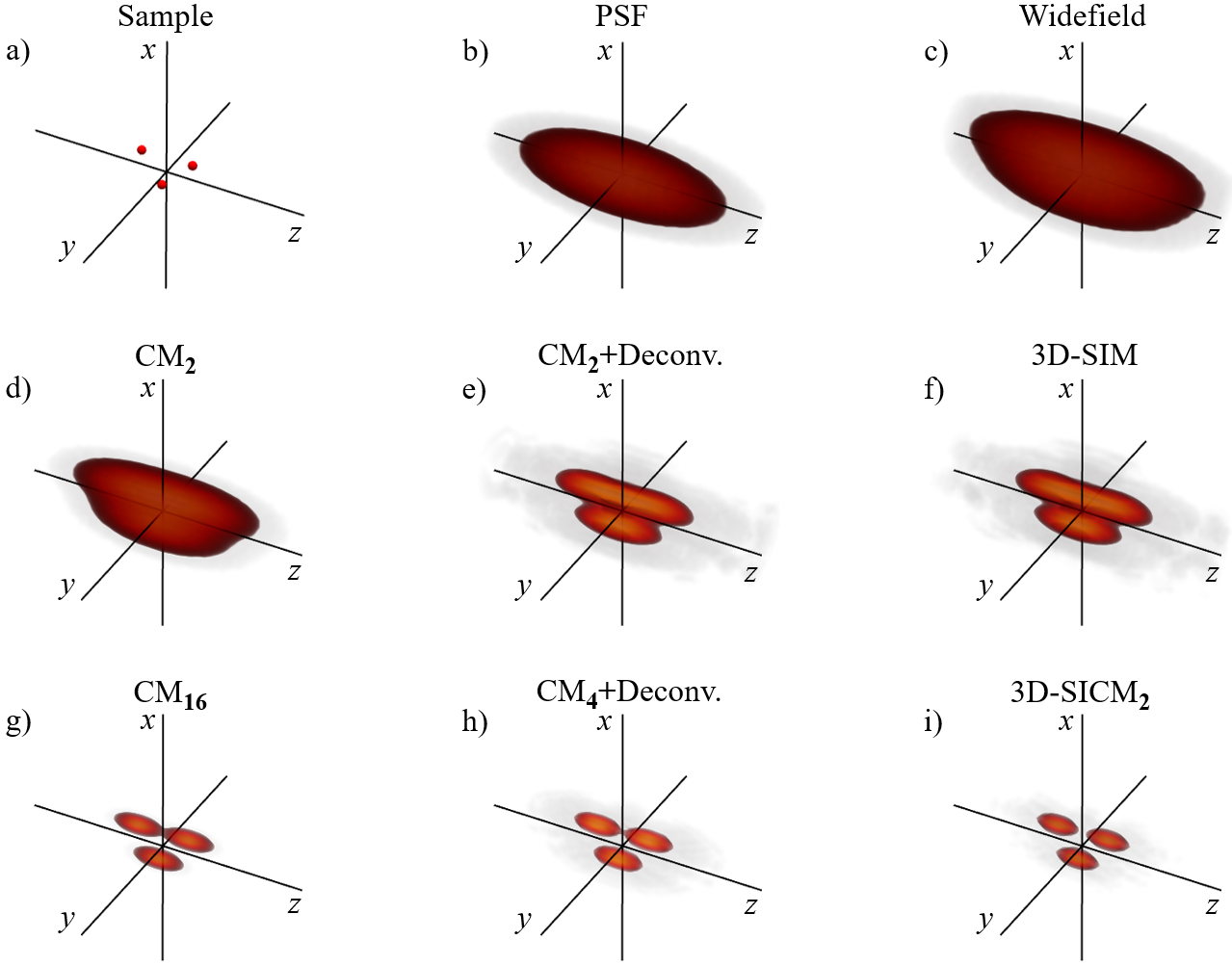}
\caption{The figure shows (a) an object  consisting of three emitters at positions $\mathbf{r}_1=(-0.16,0.16,0.05)$, $\mathbf{r}_2=(0.26,-0.26,0.57)$ and $\mathbf{r}_3=(0.26,-0.26,0.68)$ (in units of $\Delta \rho_\text{min}$) and (b) the 3D PSF utilized in the simulation. The images (c) - (i) are obtained by the methods (c) widefield microscopy, (d) second-order ICM, (e) second-order ICM + Deconvolution, (f) 3D-SIM, (g) $16^\text{th}$-order ICM, (h) fourth-order ICM + Deconvolution, and (i) second-order 3D-SI-ICM \cite{Classen2018}.}
\label{fig:3D-3}%
\end{figure} 

To remain within the linear regime either plasmonic illumination can come to the rescue \cite{Ponsetto2017}, or as very recently proposed, the analysis of intensity correlations \cite{Classen2017a}. Only recently it was realized that SIM and ICM can fruitfully be combined to enhance the lateral resolution of each technique further \cite{Classen2017a}. The same approach holds true for the axial resolution when using 3D-SIM \cite{Gustafsson2008}, thus enabling full 3D deep subwavelength resolutions, merely through application of linear optics \cite{Classen2018}. In SI-ICM the structured illumination encodes information from outside the original OTF support and the correlation analysis raises all signals to the $m^\text{th}$ power. In mathematical terms the outlined procedure corresponds to a combination of Eqs.\,(\ref{eq:sim1}) and (\ref{eq:icm1}) which results in
\begin{equation}
\text{SI-ICM}_m(\mathbf{r}) =  \sum_{i=1}^N  h(\mathbf{r}-\mathbf{r}_i)^m  \times I_{\text{str}}(\mathbf{r}_i,\alpha,\varphi)^m  \, .
\label{eq:si-icm1}
\end{equation}
Now, higher harmonics up to $\cos (m \mathbf{k}_0 \mathbf{r})$ with frequency shifts $\pm m \mathbf{k}_0$ arise, and the individual OTFs $H_m(\mathbf{k}) = FT\{ h(\mathbf{r})^m\}$ is enlarged by the factor $m$ (when including deconvolution). For correlation order $m$ the total resolution enhancement thus reaches values of $m+m = 2m$. To verify the theoretical predictions a basic simulation was conducted. The results are shown in Fig.\,\ref{fig:3D-3} and are in good agreement with the theory.

Further enhancements of the axial resolution can be achieved by combining SI-ICM with the double-objective 3D-SIM technique known as $\text{I}^5$S \cite{Shao2008}. The three added coherent beams from the second objective lead to very fast axial modulations and the axial OTF would be enlarged as in 4Pi- and $\text{I}^5$-microscopy \cite{Hell1994a,Gustafsson1999}. Another promising route may be to combine SI-ICM with plasmonic SIM techniques \cite{Zeng2014,Ponsetto2017}. Even though these techniques are limited to 2D, they allow for spatial frequencies $\mathbf{k}_{0} > \mathbf{k}_\text{max}$ of the standing wave pattern. In linear plasmonic SIM $\mathbf{k}_{0} = 2\mathbf{k}_\text{max}$ should not be exceeded to prevent gaps in the OTF support \cite{Ponsetto2017}. SI-ICM, however, would highly benefit from spatial frequencies $\mathbf{k}_0 > 2 \mathbf{k}_\text{max}$ since the enlarged OTF $H_m(\mathbf{k})$ prevents an early formation of gaps and the higher harmonics $\cos (m \mathbf{k}_0 \mathbf{r})$ would reach out to very far.

Finally, we point out that a first experiment that relies on the SI-ICM principle, while in a confocal microscopy setting, was recently demonstrated by Tenne \textit{et al.} \cite{Tenne2019}. In the paper the authors combine image scanning microscopy (which can be regarded as a confocal SIM variant \cite{Strohl2016}) with the evaluation of quantum correlations by making use of antibunching of individual quantum dots. While their setting is different from the widefield microscopy setup discussed here it delivers a first cornerstone towards real applications.

\section{Conclusion}

In this review we illuminated the original derivation of the theory of partial coherence by Wolf. Thereafter we discussed a wide range of applications where his theory plays a crucial role today, and in general how Wolf's contributions have been transformative to the field of optics. This involves applications of laser field propagation through turbulent atmosphere, optical image formation, medical diagnostics, optical coherence tomography, speckle imaging, superresolution microscopy, and studies of disorder via intensity-intensity correlation measurements. While most aspects have been examined from a classical point of view, the few detours to the realm of quantum optical aspects and techniques equally show the importance of the theory of partial coherence for these fields.

\begin{figure}[t]%
\centering
\includegraphics[width=0.495 \linewidth]{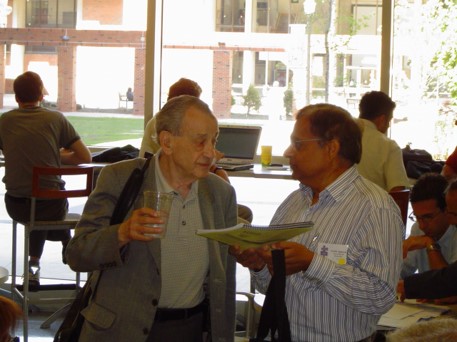}
\includegraphics[width=0.495 \linewidth]{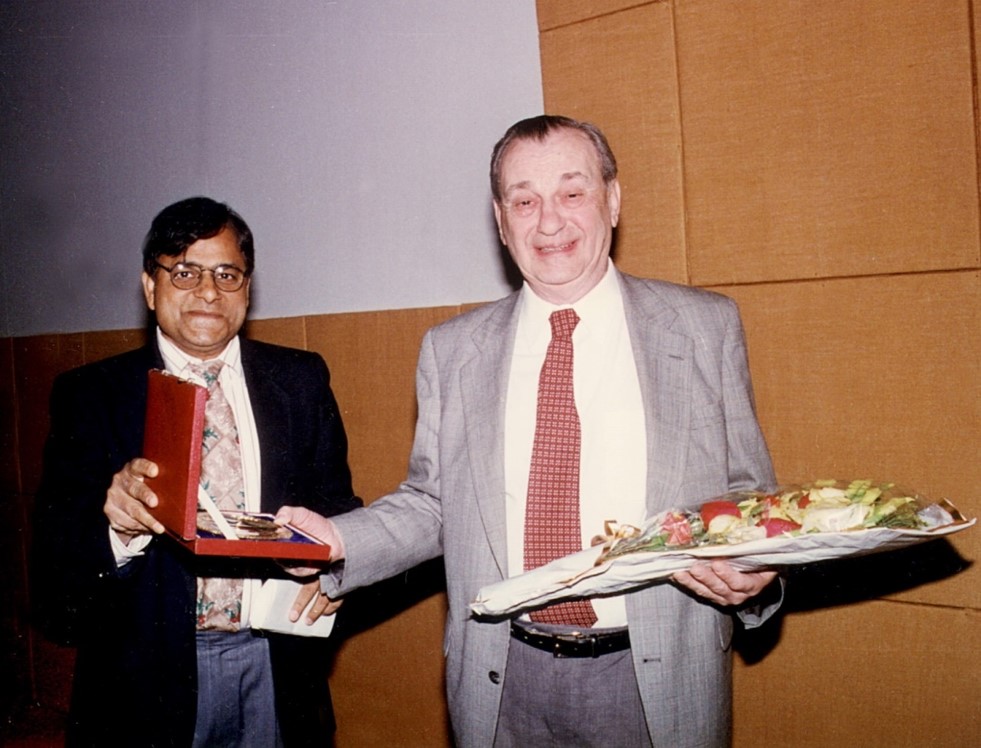}
\caption{Emil Wolf with Girish Agarwal. Left: Wolf Listening carefully. Right: Happy Wolf before Lecture at Physical Research Laboratory in India}
\label{fig:Emil}%
\end{figure}

\textit{A Personal Note from G. S. Agarwal:} I was privileged to be a graduate student of Emil in the late sixties and later a collaborator on many important issues in classical optics. More than a collaborator I learnt something new every time I met Emil over a period of 50 years. In fact whenever I visited him, he would tell me that he had several problems for me. He would have a stack of colored folders each with a different problem. I always admired his originality (Fig.~\ref{fig:Emil}). He would look into issues which have become our accepted knowledge and would come up with exceptional questions and new insights. It was through my association with Emil that I started admiring the beauty and strength of classical optics. He especially got me interested in the scattering of electromagnetic waves, a subject which I still find useful in my research today.

A. C. acknowledges support from the Alexander von Humboldt Foundation in the framework of a Feodor Lynen Research Fellowship. We acknowledge support from the Welch Foundation (Award number: A-1943-20180324). We further would like to thank Joachim von Zanthier for the long standing collaboration on higher-order intensity correlations.

\bibliography{refs_wolf}{}

\end{document}